\documentclass[12pt]{iopart}

\def \thetitle {Orbital Josephson effect and interactions in driven atom condensates on a ring}

\usepackage{graphicx}
\usepackage{setstack}
\usepackage{iopams}
\usepackage[usenames]{color}
\definecolor{extlinkcolor}{rgb}{0.7,0.15,0.15}
\usepackage[
	citecolor=blue,           % I don't like the default green for citations
	urlcolor=extlinkcolor,    % a much more decent color than cyan or magenta.
	colorlinks=true,          % colored links instead of boxes around them
	breaklinks=true,
	pdftitle={\thetitle},
	pdfsubject={Orbital Josephson Effect}
	]{hyperref}

\definecolor{RefereeItem}{rgb}{0.5,0.5,0.5}

\begin{document}

\title{Orbital Josephson effect and interactions in driven atom condensates on a ring}

\author{M. Heimsoth$^{1,2}$, C. E. Creffield$^{1}$, L. D. Carr$^{2,3}$, F. Sols$^1$}

\address{$^1$ Departamento de F\'isica de Materiales, Universidad Complutense de Madrid, 28040 Madrid, Spain}
\address{$^2$ Department of Physics, Colorado School of Mines, Golden, Colorado 80401, USA}
\address{$^3$ Physikalisches Institut, Universit\"at Heidelberg, Philosophenweg 12, 69120 Heidelberg, Germany}

\date{\today}

% 1: Tunneling of BECs, 2: Dynamic properties of BECs
% \submitto{\NJP}

\begin{abstract}
In a system of ac-driven condensed bosons we study a new type of Josephson effect occurring between states sharing the same region of space and the same internal atom structure. We first develop a technique to calculate the long time dynamics of a driven interacting many-body system. For resonant frequencies, this dynamics can be shown to derive from an effective time-independent Hamiltonian which is expressed in terms of standard creation and annihilation operators. Within the subspace of resonant states, and if the undriven states are plane waves, a locally repulsive interaction between bosons translates into an effective attraction. We apply the method to study the effect of interactions on the coherent ratchet current of an asymmetrically driven boson system. We find a wealth of dynamical regimes which includes Rabi oscillations, self-trapping, and chaotic behavior.
In the latter case, a full many-body calculation deviates from the mean-field results by predicting large quantum fluctuations of the relative particle number.
% In the latter case, a full many-body calculation deviates from the mean-field results by predicting macroscopic entanglement.
\end{abstract}

\maketitle

\section{Introduction}
The Josephson effect (JE) is a fundamental quantum phenomenon which reflects the coherent occupation of two or a few single-particle states by a macroscopic number of bosons. It was first predicted~\cite{josephson_possible_1962} and observed~\cite{experimental_JE} for superconductors, and seen later in superfluids~\cite{Varoquaux_Observation}. In Bose-Einstein condensates (BECs) one may refer to the internal or external JE, depending on whether the two coherently connected states occupy the same region of space with different internal atomic states, or two different regions with the same spin state. Both the internal~\cite{Hall_Measurement_1998,Zibold_Classical_Bifurcation_2010} and the external~\cite{albiez_direct_2005} JE have been observed. The internal JE holds the promise of generating highly entangled quantum many-body systems~\cite{Micheli_MPEntanglement_2003}. Here we report on a new type of JE which involves the macroscopic coherent occupation of Floquet states in ac-driven BECs. This {\it orbital Josephson effect} is properly neither external nor internal, since the connected Floquet states occupy the same region of space with the same internal atom structure, their sole difference residing in their orbital state.

The Josephson effect requires interactions in order to display truly collective behavior associated with the macroscopic occupation of two states~\cite{Zapata_Josephson_1998,sols1998}. In the absence of interactions, the resulting Rabi dynamics is merely an amplification of the dynamics undergone by a single atom~\cite{Hall_Measurement_1998,sols1998}. The role of interactions in ac-driven many-body systems is difficult to treat, and frequently simplifications are made such as the two-mode approximation~\cite{holthaus}, mean field theory~\cite{denisov,cec,wuester_dynamic_tunneling}, and its first~\cite{castin_instability_1997,castin_low_1998} and second~\cite{gardiner_number_2007,billam_coherence_2012} order corrections, or the use of an effective description~\cite{Ecka05-Cref06} in terms of a static many-body Hamiltonian with renormalized parameters. This latter approach is valid when the driving frequency is the dominant energy scale. Alternatively such systems can be numerically studied by the exact simulation of small clusters \cite{Ecka05-Cref06}, or by recently developed techniques such as the time-dependent density-matrix renormalization group (t-DMRG) method~\cite{schollwoeck2011}, or multiconfigurational time-dependent Hartree for bosons (MCTDHB)~\cite{MCTDBH}, both of which are able to treat larger systems. Here we develop a description of the coarse-grained dynamics of the quantum field operator in an ac-driven many-boson system, thus going beyond a mean-field treatment and its first corrections. We find that, in the case of resonantly connected Floquet states, the long-time evolution can be described by conventional quantum dynamics, by which we mean one deriving from an effective interacting Hamiltonian. If the driving provides the only external potential, we find that a repulsive interaction in real space translates into an attractive interaction in the truncated Hilbert space of a discrete number of resonantly connected Floquet states. We apply these findings to understand in depth the role of interactions in asymmetrically driven BECs which exhibit the quantum ratchet effect~\cite{denisov,cec,mh,Santos_Tailored}, together with a wealth of dynamic regimes.

% \section{$\mathbf{(t,t')}$ formalism for field operators}
\section{$\mathbf{(t,t')}$ formalism in second quantization}
\begin{figure}[t]
\centering
\includegraphics[width=0.6\textwidth]{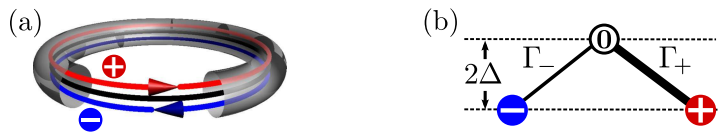}
\caption{{\it Schematics of the orbital Josephson effect}. (a)~Resonant driving induces coupling between the zero-momentum and two opposite momentum eigenmodes. The resulting Josephson link between these orbitals, with Rabi frequencies $2\Gamma_{\pm}$, has a negative effective interaction energy. (b)~A possible departure from exact resonance lifts the degeneracy of the quasi-energies.}
\label{fig: schematic}
\end{figure}

% \section{(t,t\textasciiacute) Formalism for Field Operators}
Our starting point is the equation of motion for the field operator of a system of interacting bosons,
\begin{equation}\label{eqn: Heisenberg equation for field operators}
i\partial _{t}\hat{\psi}(x,t)
 = \big[H(x,t)+\lambda \hat{\psi}^{\dag}\hat{\psi}\big]\hat{\psi}(x,t) ,
\end{equation}
where $H(x,t)=-\frac{1}{2}\partial_{xx}+V(x,t)$ includes
a time-dependent potential, and $\lambda $ is the effective strength of the
contact interaction in a quasi-one-dimensional ring of radius $R$.
We set $\hbar=1$ and
measure all energies and frequencies in units of $\hbar^2/MR^2$, with $M$ the atomic mass. Furthermore, our unit of length is $R$, so that the circumference of the ring is $2\pi$.
Such rings, as schematically depicted in Fig.~\ref{fig: schematic}a, have been proposed and experimentally realized -- see e.g. Refs.~\cite{phillips_torus_2011,tori}, among many other ring experiments. A possible particle flow can be observed via standard time-of-flight measurement techniques~\cite{phillips_torus_2011}. The field operator $\hat{\psi}(x,t)$ obeys periodic boundary
conditions in $x$, and its algebraic structure is given by the standard
equal-time bosonic commutation relations: $[\hat{\psi}(x_1,t),\hat{\psi}%
^{\dag}(x_2,t)]=\delta(x_1-x_2)$ and $[\hat{\psi}(x_1,t),\hat{\psi}(x_2,t)]=0$.

In order to study the long-term dynamics of the many-body system, we generalize the $(t,t')$~formalism to treat second-quantized operators. This formalism was originally developed for single particles~\cite{tt'}, and later adapted to treat helium~\cite{helium_tt'} always within first quantization.
% In the following, we will present an extension of the $(t,t')$~formalism to the dynamic of field operators in Heisenberg picture.
We do this extension by introducing an additional parameter $t'$ which the field operators depend on, and impose periodic boundary conditions in $t'$ with period~$T$
\begin{equation}
 \label{eqn: periodic boundary conditions in t'}
 \hat{\psi}(x,t'+T;t)=\hat{\psi}(x,t';t) 
\end{equation}
We define $\hat{\psi}(x,t';t)$ as the solution of the equation of motion
\begin{equation}
  \label{generalized-Heisenberg-eqn}
i\partial_t\hat{\psi}(x,t';t)
  =\big[H(x,t')-i\partial_{t'}
   +\lambda \hat{\psi}^\dag\hat{\psi}\big]\hat{\psi}(x,t';t)\,,
\end{equation}
together with the initial condition
\begin{equation}
 \label{eqn: initial conditions for extended field operators}
 \hat{\psi}(x,t';0)\equiv\hat{\psi}(x,0)\,.
\end{equation}
% 
% The equation of motion and initial conditions for the creation operators are given by the hermitian adjoint of Eqs.~(\ref{generalized-Heisenberg-eqn})-(\ref{eqn: initial conditions for extended field operators}).
The periodic boundary conditions in $t'$, which are imposed initially via Eq.~(\ref{eqn: initial conditions for extended field operators}), are guaranteed to be preserved over time~$t$ due to the periodicity of $H(x,t')$.

% The equation of motion~(\ref{generalized-Heisenberg-eqn}) does not lead outside the space of operators acting on Fock space.
We note that the field operator $\hat{\psi}(x,t';t)$ acts on the same Fock space as the Heisenberg field operators $\hat{\psi}(x,t)$, for all $t'$ and all times $t$.

We may choose $t'$ to be restricted to be a function of $t$ given by $t'_\tau(t)=t+\tau$, with~$\tau\in[0,T]$.
% \footnote{Note that due the periodicity in~$t'$, one could also take $t'_\tau=(t+\tau)\mod T$ and allow for values of $\tau$ out of the interval $[0,T]$.}
Then the resulting operator $\hat{\psi}_\tau(x,t)\equiv\hat{\psi}(x,t'_\tau;t)=\hat{\psi}(x,t+\tau;t)$ is the solution of the Heisenberg equation~(\ref{eqn: Heisenberg equation for field operators}), but with a shifted switching $H_\tau(x,t)=H(x,t+\tau)$ and initial condition $\hat{\psi}_\tau(x,0)=\hat{\psi}(x,0)$. This can be shown by calculating the total time derivative along the line~$t'=t'_\tau(t)$:
\begin{eqnarray}
\label{eqn: prove extended formalism}
%  i\frac{\partial}{\partial t}\hat{\psi}_\tau(x,t)
 i\partial_t\hat{\psi}_\tau(x,t)
 &=&i\frac{d}{dt}\hat{\psi}(x,t'_\tau;t)\Big|_{t'_\tau=t+\tau}
 =\Bigg[i\frac{\partial\hat{\psi}}{\partial t'_\tau}\frac{dt'_\tau}{dt}
 +i\frac{\partial\hat{\psi}}{\partial t}\Bigg]_{t'_\tau=t+\tau} \nonumber\\
 &=&[H(x,t+\tau)+\lambda\hat{\psi}^\dag\hat{\psi}]\hat{\psi},
\end{eqnarray}
where we have used Eq.~(\ref{generalized-Heisenberg-eqn}) and the identity $dt'_\tau/dt=1$. In other words, $\hat{\psi}(x,t'_\tau;t)=\hat{\psi}_\tau(x,t)$ is but a solution of the Heisenberg equation of motion for a Hamiltonian shifted in time by $\tau$ with the initial condition $\hat{\psi}_\tau(x,0)=\hat{\psi}(x,0)$.
In particular $t'=t\;(\tau=0)$ gives the solution to Eq.~(\ref{eqn: Heisenberg equation for field operators}).
% Since one can find a $\tau$ for any $t'$, such that $t'=t'_\tau(t)$, we obtain the commutation relations
This implies the following commutation relations for the solutions of Eqs.~(\ref{generalized-Heisenberg-eqn})-(\ref{eqn: initial conditions for extended field operators})
\begin{equation}
 \label{eqn: non-standard commutation relation}
 \eqalign{[\hat{\psi}(x_{1},t';t),\hat{\psi}^\dag(x_2,t';t)] =\delta(x_{1}-x_{2}), \cr
 [\hat{\psi}(x_{1},t';t),\hat{\psi}(x_2,t';t)]
%  =[\hat{\psi}^\dag(x_{1},t';t),\hat{\psi}^\dag(x_2,t';t)]
 =0~.}
\end{equation}

The advantage of the $(t,t')$~approach is that it provides a natural separation of timescales. The $t'$ coordinate describes the behavior of the system on timescales shorter than the driving period $T$, while the long-time dynamics is described by $t$ -- see Eqs.~(\ref{unperturbed-stationary})-(\ref{unperturbed-Floquet-state}) below. Furthermore, the originally time-dependent problem [see Eq.~(\ref{eqn: Heisenberg equation for field operators})] is mapped to a formally time-independent one [see Eq.~(\ref{generalized-Heisenberg-eqn})], which opens the possibility of using resolution methods developed for such type of problems~\cite{mh,tt',helium_tt'}.

The field-operator extension of the $(t,t')$-formalism which we have presented is rather general. It is neither restricted to bosonic systems, nor to the ring geometry we study here. It can be straightforwardly translated to the case of fermionic or mixed systems, as well as to other trap-geometries.

Due to the periodic boundary conditions in $t'$, it is natural to express the $t'$-dependence of $\hat{\psi}(x,t';t)$ via a Fourier decomposition, and we may approximate the dynamics by incorporating only a few modes.
% Since the $t$-dependence lacks periodicity for both $\hat{\psi}(x,t)$ and $\hat{\psi}(x,t';t)$, a Fourier decomposition in the $t$-dependence would be rather unnatural.
% Question: Fernando wants to skip this, but Lincoln had asked for it, but only because Fernando had doubts .... uff.

To include the effect of a weak time-periodic perturbation, $%
V(x,t)=V(x,t+T)$, it is convenient to work in the representation of
unperturbed ($V=0$) Floquet states. The first-quantized version of (\ref{generalized-Heisenberg-eqn})
was studied in Ref.~\cite{mh} in
the absence of interactions.
The unperturbed stationary states evolve as
\begin{equation}
\psi _{\ell m}(x,t^{\prime };t)=\phi _{\ell m}(x,t^{\prime })\exp (-i\varepsilon
_{\ell m}^{0}t)\,,  \label{unperturbed-stationary}
\end{equation}
where $\varepsilon_{\ell m}^{0}=\frac{1}{2}\ell^2-m\omega $ and $\ell,m$ are integers labeling the Fourier modes
\begin{equation}
\phi _{\ell m}(x,t^{\prime })
%=\langle x,t^{\prime }|\ell m\rangle
=\frac{1}{\sqrt{2\pi}}\exp
(i\ell x-im\omega t^{\prime })\,,  \label{unperturbed-Floquet-state}
\end{equation}
which are the unperturbed Floquet states with quasi-energy
$\varepsilon_{\ell m}^{0}$, with $\omega=2\pi/T$. The operators in this representation,
\begin{equation}
\hat{a}_{\ell m}(t)=\frac{1}{T}\int \!\!\!\int\!dx~dt^{\prime }\phi
_{\ell m}^{\ast }(x,t^{\prime })\hat{\psi}(x,t^{\prime };t)~,
\label{projection}
\end{equation}
satisfy the equation of motion
\begin{eqnarray}
i\partial _{t}\hat{a}_{\ell m}
 &=&\varepsilon _{\ell m}^{0}\hat{a}_{\ell m}
  + \sum_{\ell^{\prime }m^{\prime }}
           V_{\ell m,\ell^{\prime }m^{\prime }}\hat{a}_{\ell^{\prime }m^{\prime }}
  \label{lm-eqn-motion} \\
&&+\frac{\lambda }{2\pi}
      \sum_{\ell^{\prime }m^{\prime }\ell^{\prime \prime}m^{\prime \prime }}
         \hat{a}_{\ell^{\prime }m^{\prime }}^{\dag }
         \hat{a}_{\ell^{\prime \prime }m^{\prime \prime }}
         \hat{a}_{\ell^{\prime }-\ell^{\prime \prime}+\ell,m^{\prime }-m^{\prime \prime }+m}
  \, , \nonumber
\end{eqnarray}
and the commutation relations~(\ref{eqn: non-standard commutation relation}) translate to
\begin{equation}
 \label{unconv-commut}
 \eqalign{
  \sum_{m'}[\hat{a}_{\ell,m'+m}(t),\hat{a}^{\dag}_{\ell' m'}(t)]
  =\delta_{\ell\ell'}\delta_{m0}\,, \cr
  \sum_{m'}[\hat{a}_{\ell,m'-m}(t),\hat{a}_{\ell' m'}(t)]
%   =\sum_{m'}[\hat{a}^{\dag}_{\ell,m'-m}(t),\hat{a}^{\dag}_{\ell',-m'}(t)]
  =0\,,\mbox{for all }m.
  }
\end{equation}
If one were to view $m$ as just an additional standard (orbital or spin) quantum number, these commutation relations would appear unconventional (very much as Eqs.~(\ref{eqn: non-standard commutation relation}) would look unconventional if $t'$ were regarded as an extra space variable). However, Eqs.~(\ref{unconv-commut}) follow naturally from the Fourier transformation of Eqs.~(\ref{eqn: non-standard commutation relation}).

Here $V_{lm,l'm'}$ is the matrix element of the driving operator between two unperturbed states, as given in Eq.~(\ref{unperturbed-Floquet-state}).
When the driving frequency is such that the system is at or close to resonance,
only a few states (all with the same or similar value of
$\varepsilon _{\ell m}^{0}$)\ are relevant~\cite{denisov,cec,mh}. Importantly, in that subspace the index $m$ is
uniquely determined by $\ell$. Thus within that truncated space we can drop the index $m$.
As a result, Eqs.~(\ref{unconv-commut}) become equivalent to the standard bosonic commutation relations
$[\hat{a}_{\ell}(t),\hat{a}_{\ell^{\prime }}^{\dag }(t)]=\delta_{\ell\ell^{\prime }}$ and $[\hat{a}_{\ell}(t),\hat{a}_{\ell^{\prime }}(t)]=0$.

% , with the quasi energy defined as $\varepsilon_\ell^0=\frac{1}{2}\ell^2\mathrm{mod}\omega$.

\section{Resonant driving}
If we calculate the effective
matrix elements connecting the various resonant unperturbed Floquet
states, we are left with a conventional few-mode boson problem whose
dynamics, given by Eq.~(\ref{lm-eqn-motion}), can be studied using established
techniques. The resulting dynamics
between Floquet states reflects the coarse-grained, long-time dynamics of
the true quantum state evolution.

Within the degenerate (or almost degenerate) subspace where conventional commutation relations apply, Eq.~(\ref{lm-eqn-motion}) can be viewed
as a Heisenberg equation deriving from the interacting many-body Hamiltonian
\begin{equation}
\hat{H}
  = \sum_{\ell}\varepsilon_{\ell}^{0}\hat{n}_{\ell}
   + \sum_{\ell\ell'}\Gamma_{\ell\ell'}\hat{a}_{\ell}^{\dag}\hat{a}_{\ell'}
   + \frac{\lambda}{4\pi}
        \sum_{\ell\ell'}\sum_{\ell''}
                \hat{a}_{\ell+\ell''}^{\dag}
                \hat{a}_{\ell'-\ell''}^{\dag}
                \hat{a}_{\ell}\hat{a}_{\ell'}~,
  \label{H-pair}
\end{equation}
where $\varepsilon_{\ell}^{0}=\frac{1}{2}\ell^2 \bmod{\omega}$, $\Gamma_{\ell\ell'}$ may allow for second-order processes between resonant states
mediated by a non-resonant state~\cite{mh},  $\hat{n}_{\ell}=\hat{a}_{\ell}^{\dag }\hat{a}_{\ell}$ is the occupation of
state $\ell$, and $\ell''$ can only take values $0$ or $\ell'\!\!-\!\ell$ if $\ell\neq\ell'$, and $0$ if $\ell=\ell'$.

We remark that the derivation of an effective Hamiltonian does not require~$\ell$ to describe a plane wave. In such a case, the interaction term in Eq.~(\ref{H-pair}) would contain a sum over four orbital indices, and the matrix elements would depend on all four involved orbitals.
% % except for the simplifying selection rules there implicit. (which make the interaction matrix elements independent of the involved quantum number).
% However the advantages of a plane wave is that the matrix elements of the interaction simplify remarkably.

The conservation of total particle number $\sum_\ell \hat{n}_\ell =N$,
% \begin{equation}
% \sum_\ell \hat{n}_\ell |\Psi\rangle = \hat{N} |\Psi\rangle = N |\Psi\rangle
% \end{equation}
% for all system states $|\Psi\rangle$,
% allows us to write
together with the identity
\begin{equation}\label{eqn: extracting diagonal terms}
 \sum_{\ell\neq\ell'}\hat{n}_{\ell}\hat{n}_{\ell'}=N^2-\sum_\ell n_\ell^2\, ,
\end{equation}
%
% \begin{equation}\label{eqn: extracting diagonal terms}
% \sum_{\ell \neq \ell'}
%    = \bigg(\sum_{\ell}  \hat{n}_{\ell}\bigg)
%      \bigg(\sum_{\ell'} \hat{n}_{\ell'}\bigg)
%     - \sum_\ell \hat{n}_{\ell}^2 = \hat{N}^2 - \sum_\ell n_\ell^2 \, ,
% \end{equation}
%
leads to the interesting result that,
up to a constant $(2N^2 - N) \lambda/4 \pi$, the Hamiltonian~(\ref{H-pair}) is equivalent to
\begin{equation}
\hat{H}
 = \sum_{\ell}\varepsilon _{\ell}^{0}\hat{n}_{\ell}
 + \sum_{\ell\ell'}\Gamma_{\ell\ell'}\hat{a}_{\ell}^{\dag}\hat{a}_{\ell'}
   -\frac{\lambda }{4\pi}\sum_{\ell}\hat{n}_{\ell}^{2}~.
\label{H-attractive}
\end{equation}
Thus a {\em repulsive} interaction in real space translates into an
{\em attractive} interaction in (angular) momentum space. The resulting energy gain in the macroscopic occupation of a
single state in gases with repulsive interactions underlies the stability of
Bose-Einstein condensation, as discussed in Ref.~\cite{tony_book}.

To derive (\ref{H-attractive}), we have used the selection rules (stemming from the simultaneous conservation of momentum and quasi-energy) which are implicit in Eqs.~(\ref{lm-eqn-motion})-(\ref{H-pair}). Interestingly, in those dynamical regimes where the truncated resonant space is effectively reduced to two modes, Eq. (\ref{H-attractive}) applies for orbitals of arbitrary shape (not necessarily plane waves), but with an interaction strength that is state dependent.

When a few modes are macroscopically occupied, Eq. (\ref{H-pair})
describes a Josephson-type link between states $\ell\ell'$ with Rabi
frequency $2\Gamma_{\ell\ell'}$. Remarkably, this Josephson link takes
place between atom states with the same internal state and occupying the
same region of space. Thus it is not appropriate to describe it as an internal
or external JE. The Hamiltonian  (\ref{H-attractive}) describes what can be
termed the orbital Josephson effect, which here occurs between
resonant Floquet states in ac-driven Bose condensates.

\section{Application to the quantum ratchet}
As a particular application of the method developed here to treat the many-body problem in driven systems, we consider a BEC subject to the asymmetric driving
\begin{equation}
V(x,t)=K[\sin (kx)+\alpha \sin (2kx+\varphi )][\sin (\omega t)+\beta \sin
(2\omega t)]~,
\end{equation}
which has been numerically studied~\cite{cec} and experimentally implemented by Salger et al.~\cite{bonn} on an extended optical lattice. Since the driving in Ref.~\cite{bonn} conserves quasi momentum, the here considered periodic boundary conditions can be related to the experimentally more convenient lattice system.
For $\varphi\neq\frac{1}{2}\pi,\frac{3}{2}\pi$ it yields a coherent quantum ratchet provided that both $\alpha$
and $\beta $ are nonzero.

We consider a system initially prepared in $|\ell\rangle=|0\rangle$.
It was analytically shown in Ref.~\cite{mh}
that, if we drive it with resonant frequency $\omega=1$ and small amplitude $K$, the subsequent evolution will mix
the initial state only with the states $|\ell\rangle =|\pm2\rangle$.
These three states
all satisfy the resonance condition $\varepsilon _{\ell}^{0}=0$, and we label them with the indices  $\pm ,0$. The resulting three-level system Hamiltonian is
\begin{equation} \label{truncated}
\hat{H}_{\mathrm{3LS}} =
%  \phantom{-}
  \Gamma_{+}\hat{a}_{+}^{\dag}\hat{a}_{0}
       + \Gamma_{-}\hat{a}_{-}^{\dag}\hat{a}_{0}
       + \mathrm{H.c.}
%   \nonumber\\
    - \frac{\lambda }{4\pi }\sum_{\nu}\,\hat{n}_{\nu }^{2}
    - 2\Delta (\hat{n}_{+}+\hat{n}_{-}) ,
\end{equation}
where $\nu$ takes values $\pm,0$,
and $\Delta =\omega-1$ accounts for a
possible small detuning that shifts the quasi-energies in Eq.~(\ref{H-attractive}) to $\varepsilon _{\ell}^{0}=-(\ell^2/2)\Delta$. The resulting level structure is schematically depicted in Fig.~\ref{fig: schematic}b. Near resonance the tunneling parameters $\Gamma_{\pm }$
can be calculated analytically~\cite{mh}. For $\Delta \ll 1$, one finds
\begin{equation}
\Gamma _{\pm }=\frac{K}{4}\Bigg(\frac{K}{2}\pm \alpha \beta e^{\pm
i\varphi }\Bigg).  \label{Gamma-pm}
\end{equation}
This result demonstrates that the quantum ratchet
current (which requires $|\Gamma _{+}|\neq |\Gamma _{-}|$) originates in the
interference between first- and second-order processes in the driving
strength~\cite{mh}.
In Eq.~(\ref{Gamma-pm}) we have neglected the possible effect of interactions on the second-order tunneling terms.
Without interactions, the initial state $|0\rangle$ couples only to a state $|a\rangle$ which is an asymmetric combination of $|\pm \rangle$, yielding an average current which is half that carried by $|a\rangle$~\cite{mh}. It has been  noticed that interactions destroy the coherent current~\cite{cec}. Below we study the role of interactions in depth and show that they yield a rich variety of dynamical regimes.

\section{Numerical results}
\label{sec: Numerical results}
We employ three different approximations to investigate the many-body problem: full numerical resolution of the time-dependent Gross-Pitaevskii equation (FGP) using a fourth-order Runge-Kutta method with a fixed time-step; study of the GP equation in the truncated space of the three resonant states $0,\pm$ (3GP); and resolution of the many-body problem in the three-level system (3LS), via exact diagonalization of the effective interacting Hamiltonian (\ref{truncated}) with up to 40 atoms. Importantly we note that 3GP can be obtained both as a truncation of FGP or a mean-field version of 3LS. We always assume the condensate to be initially in state $0$.
\begin{figure}[t]
\centering
% \raggedleft
\includegraphics[width=0.99\textwidth]{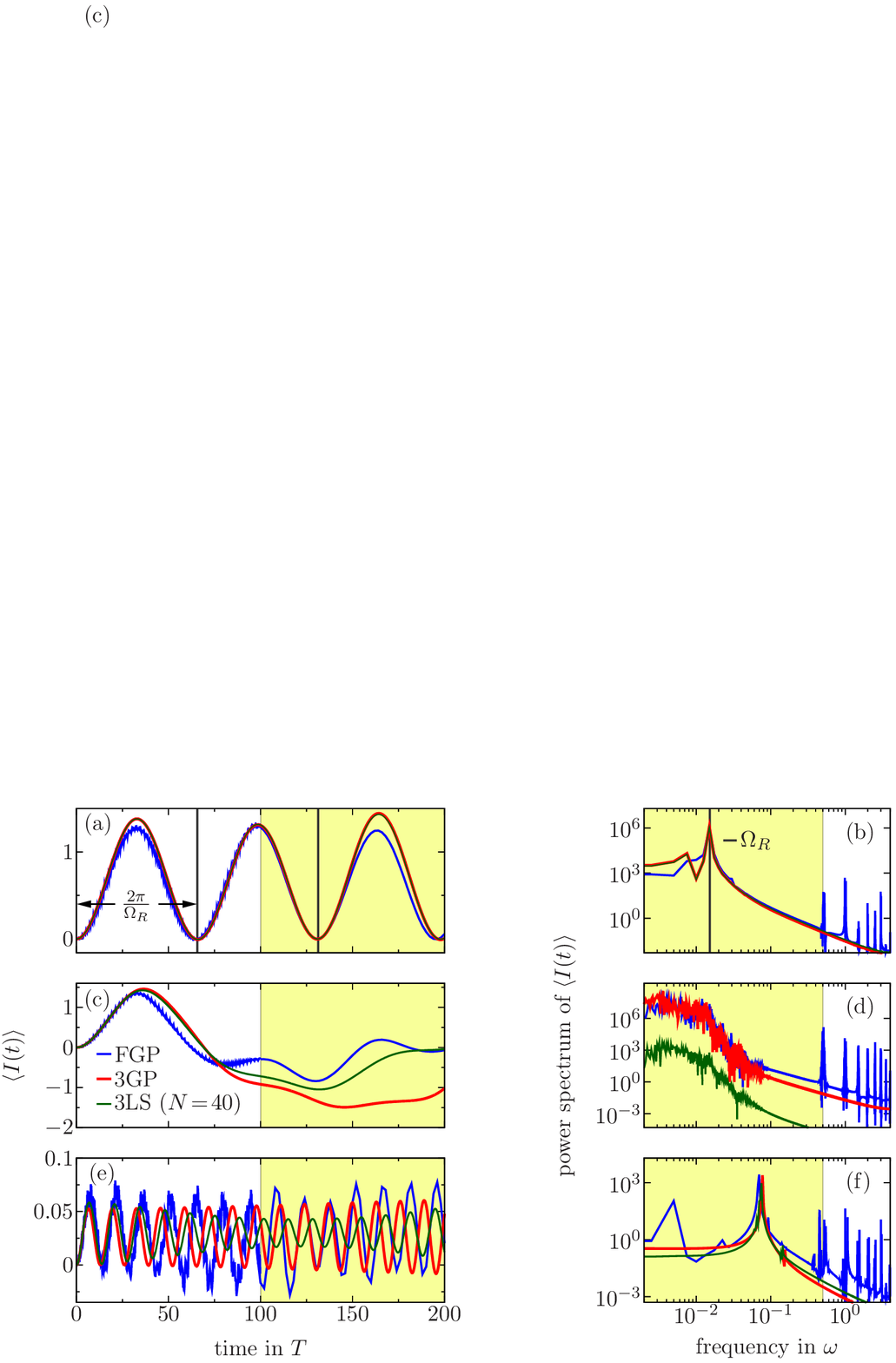}
\caption{{\it Time evolution and power spectrum of the particle current.} Three regimes are depicted: (a,b)~weak interaction strength ($g\!=\!0.01$), showing regular Rabi oscillations; (c,d)~intermediate interaction ($g\!=\!0.08$), displaying chaotic dynamics; and (e,f)~strong interaction ($g\!=\!0.5$), showing self-trapping. Parameters: $K=0.2$, $\varphi=0$, $\omega=1\;(\Delta=0)$, $\alpha=\beta=0.2$. In the shaded region we present the stroboscopic current, which erases the short time dynamics responsible for the peaks in the white region of the current spectrum. Three different calculation methods are employed: Full Gross-Pitaevskii~(FGP), three-level mean-field description~(3GP), three-level many-body description~(3LS) (see section~\ref{sec: Numerical results}). In (a,b) the green curve almost coincides with the red curve.}
\label{fig: evolution examples}
\end{figure}

\subsection{Time-dependent current}
Figure~\ref{fig: evolution examples} shows the expectation value of the ratchet current $\langle I(t)\rangle$ -- given by the mean momentum per particle -- over the first 200 driving cycles. The regimes of weak, moderate, and strong interaction [as characterized by $g\equiv\lambda(N-1)$], are all calculated with the three methods described above. For weak interactions, the two truncated-space calculations (3GP and 3LS) yield similar Rabi oscillations, both differing from FGP in that they do not display high-frequency dynamics, as revealed both in the Fourier spectrum and in the time dependence of the current. The fast dynamics of FGP disappears if the stroboscopic current is plotted, as shown in the shaded region.

For strong interactions, the system tends to remain trapped in the initial state~\footnote{In the absence of decoherence the initial state $0$ would mix with resonant states $\pm$ after a very long time.},
as may be expected since Eq.~(\ref{truncated}) has the structure of a Josephson link with {\em negative} interaction energy~\cite{kohler_oscillatory_2002}.
The three numerical methods predict differing behaviors of the resulting small current, but agree on the position and strength of the main peak in the spectrum. For intermediate interaction, discrepancies between the three methods soon appear, which suggests chaotic behavior. This is confirmed and studied further below.

Within the accuracy limited by the $\omega$-sampling, the positions of the main peaks coincide exactly in the Rabi regime, and vary by about $10\%$ in the self-trapping case.
The peak heights vary by about $6\%$ in the Rabi case, and about $9\%$ between FGP and 3GP in the the self-trapped regime.
% However for 3LS, the peaks strength is by a factor $3$ below the value predicted by both mean-field methods.
However for 3LS, the discrepancy with FGP (and 3GP) is worse (differing by a factor of $3$), which coincides with the observation that for the self-trapped dynamics one has to consider particle numbers far beyond $N=40$ in order to get an agreement between the full many-body description and the mean-field result~\cite{Milburn_1997}.

\subsection{Time-averaged current}
\begin{figure}[t]
\centering
% \raggedleft
\includegraphics[width=0.83\textwidth]{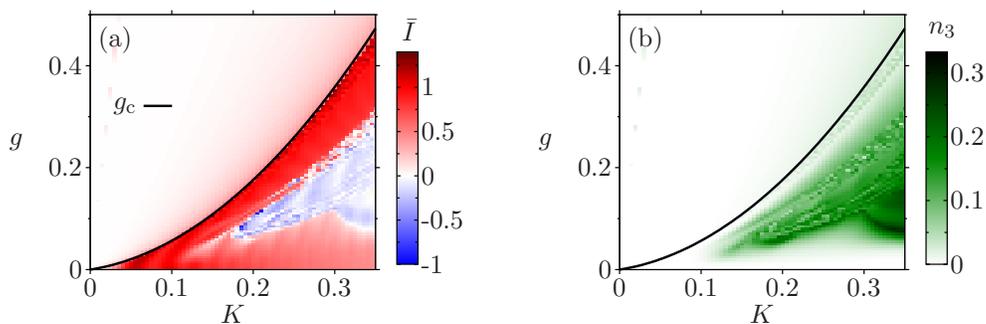}
\caption{(a)~Continuous time average of the quantum ratchet current at long times (400 driving cycles) for different amplitudes and interaction strength. (b)~Third largest eigenvalue of the time-averaged one-particle density matrix after the same interval.
%$\rho_{\mu\nu}(\tau)=\frac{1}{\tau}\int_0^{\tau}dtA_{\mu}^*(t)A_{\nu}(t)$.
Parameters: $\omega=1\;(\Delta=0)$, $\alpha=\beta=0.2$, $\varphi=0$.}
\label{fig: ratchet current and n3}
\end{figure}

Figure~\ref{fig: ratchet current and n3}a shows the continuous time average of the ratchet current
\begin{equation}
\bar{I}(t)\equiv\frac{1}{t}\int_0^t\!dt'\, \langle I(t')\rangle
\end{equation}
after $400$ cycles, computed within 3GP, as a function of $g$ and $K$ at exact resonance. We assume $|\Gamma_+|>|\Gamma_-|$; see Eq.~(\ref{Gamma-pm}). For small $K$, the perturbative scheme of~\cite{mh} applies, and at weak interaction a nonzero ratchet current results from Rabi oscillations between $|0\rangle$ and $|a\rangle$. For stronger interactions the ratchet current disappears, because the system remains in the initial state $|0\rangle$ due to the large negative energy associated with the initial macroscopic occupation of $|0\rangle$ [see Eq. (\ref{H-attractive})]. This can be viewed as a form of macroscopic quantum self-trapping~\cite{smerzi_quantum_1997}, albeit with a negative effective interaction energy. We find that just below the critical interaction strength the system oscillates between $|0\rangle$ and $|+\rangle$. This is consistent with the fact that the threshold interaction value obtained from that assumption, $g_{\mathrm{c}}=8\pi|\Gamma_+|$, agrees well with the numerical simulation (see solid curve in Fig~\ref{fig: ratchet current and n3}a). Finally, we notice the existence of a region of suppressed or weakly reversed current for stronger driving and intermediate interaction.

\subsection{Effect of detuning}
\begin{figure}[t]
\centering
% \raggedleft
\includegraphics[width=0.99\textwidth]{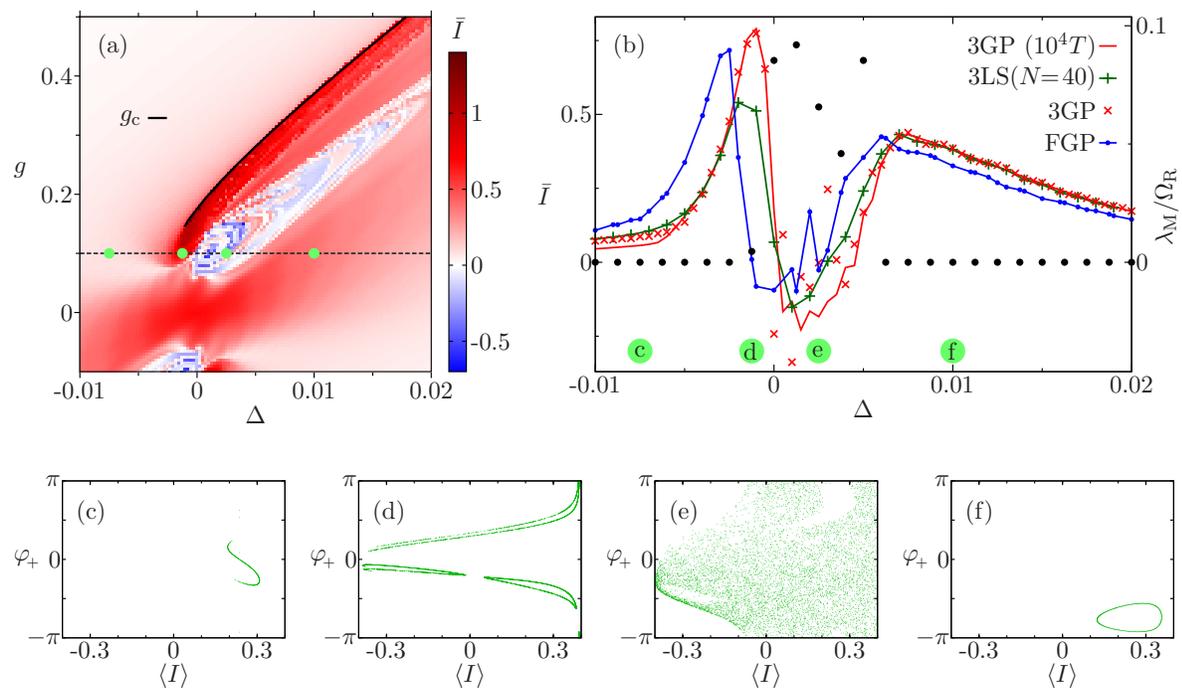}
\caption{(a)~Quantum ratchet current as a function of the interaction strength and the detuning. The behavior along the dashed line ($g=0.1$) is shown in (b)~for four different calculations (3GP is computed at two different times, $400T$ and $10^4T$). The black dots in~(b) show the maximal Lyapunov exponents and refer to the scale on the right ($\Omega_{\mathrm{R}}$ is the Rabi frequency for the non-interacting case, see Ref.~\cite{mh}). Poincar\'e sections are shown for the representative points~(c-f), obtained from points in the hyperplane $n_+\!+\!n_-\!=\!0.2$. The plotted variables are the instantaneous ratchet current $\langle I\rangle$ and the relative phase $\varphi_{+}$ between states $|0\rangle$ and $|+\rangle$. Parameters: $\alpha=\beta=0.2$, $K=0.2$, $\varphi=0$. In~(a-b) averages are performed over $400$ cycles unless otherwise indicated. }
\label{fig: detuning and sections}
\end{figure}

Figure~\ref{fig: detuning and sections}a shows the effect of departing from exact resonance. In general, the effect of $\Delta >0$ is that of shifting the behavior of current towards higher interactions. In particular we note that, starting from the white-blue region of suppressed or reversed current, interactions tend to restore the positive ratchet current. This effect can be understood if one notes that, for  positive detuning, the effective degeneracy between the three states $\pm,0$ disappears because $|0\rangle$ acquires a higher energy. Degeneracy is restored due to the massive initial occupation of $|0\rangle$, which lowers its energy via the attractive mean-field interaction in Eq.~(\ref{H-attractive}). It thus becomes clear that the effect of a positive detuning is counteracted by an increase of the effective attraction. In particular, the starting degeneracy recovered with the help of interactions permits the onset of Rabi oscillations known to be essential for the emergence of a ratchet current~\cite{mh}.

\subsection{Phase dependence and interaction symmetry}
\begin{figure}[t]
\centering\includegraphics[width=0.5\textwidth]{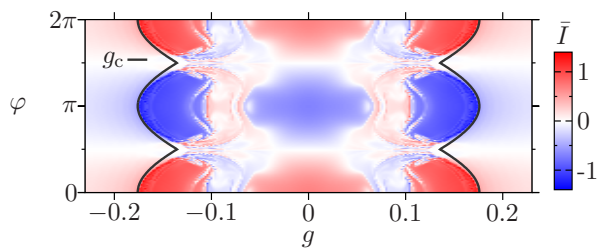}
\caption{{\it Current dependence on $\varphi$.} Ratchet current in 3GP as a function of the phase $\varphi$ and the interaction strength $g$. Parameters: $\omega=1\;(\Delta=0)$, $K=0.2$, $\alpha=\beta=0.2$. Averages are performed over $400$ cycles.}
\label{fig: current over phi and g}
\end{figure}

In Fig.~\ref{fig: current over phi and g} we show how the the ratchet current depends on the choice of initial driving phase. A change $\varphi=0 \rightarrow \pi$ corresponds to $K \rightarrow -K$ and thus to a reversal of the current. On the other hand, we note that for $\varphi=\frac{1}{2}\pi,\frac{3}{2}\pi$ the ratchet current is always zero, because parity is restored for those values of $\varphi$.
Figure~\ref{fig: current over phi and g} also shows that the long-time averaged current (but not the instantaneous one) is an even function of the interaction strength.

\subsection{Chaotic dynamics}
Figure~\ref{fig: ratchet current and n3}b shows the population $n_3$ of the third eigenstate of the time-averaged one-particle density matrix (calculated in 3GP)
\begin{equation}\label{eqn: def: time-averaged density matrix in 3GP}
 \bar{\rho}_{\mu\nu}(t)\equiv\frac{1}{t}\int_0^t dt'A^*_\mu(t')A_\nu(t')~,
\end{equation}
where $A_\mu(t)$ ($\mu=0,\pm$) are the time-dependent expansion coefficients of the condensate orbital.
Comparison with Fig.~\ref{fig: ratchet current and n3}a shows a clear correlation between the occupation of a third state and the suppression or weak reversal of the ratchet current. The dynamics of a macroscopic condensate in a three-level system is that of two coupled non-rigid pendula, which beyond the harmonic regime can be expected to be chaotic. This behavior is confirmed from the inspection of Figs.~\ref{fig: detuning and sections}b-f. In Fig.~\ref{fig: detuning and sections}b we show a plot of $\bar{I}$ as a function of $\Delta$ for $g=0.1$, obtained with the three different calculation methods described before. For the values of $\Delta$ marked with labels c-f, we show in Figs.~\ref{fig: detuning and sections}c-f the corresponding Poincar\'e cross sections. The one point which shows chaotic behavior is e, which falls in the white-blue region of Fig.~\ref{fig: detuning and sections}a. We have just noted (see Fig.~\ref{fig: ratchet current and n3}a) that such a region is correlated with the occupation of a third eigenstate and thus with the possible emergence of chaotic behavior on the coarse-grained time scale. This underlines the connection between three-level occupation, chaotic behavior, and current reduction or reversal. A calculation of the maximal Lyapunov exponent $\lambda_{\mathrm{M}}$ (given by black dots in Fig.~\ref{fig: detuning and sections}b) further confirms the chaotic character of this parameter region, as $\lambda_{\mathrm{M}}$ acquires positive finite values there. For more details, see \ref{sec: Convergence analysis}.

Figure~\ref{fig: detuning and sections}b also indicates that in the chaotic regime the convergence in time is slower than in the regular regime, as can be seen by the discrepancy between 3GP calculations performed over $400$ or $10^4$ driving periods. We emphasize that the current reversal feature in the chaotic region is not just a transient but reflects a lasting behavior; we have checked it up to 10$^5$ periods in some cases.
Finally, we note that all calculations performed in the FGP approximation indicate that, for large $K$ and $g$ close to the self-trapping transition $g_{\mathrm{c}}$ (beyond which the system essentially remains in $|0\rangle$), the behavior undergoes a significant influence from states beyond the three-state basis $0,\pm$ (not shown).

\subsection{Two-particle correlation functions}
\begin{figure}[t]
\centering
% \raggedleft
\includegraphics[width=0.99\textwidth]{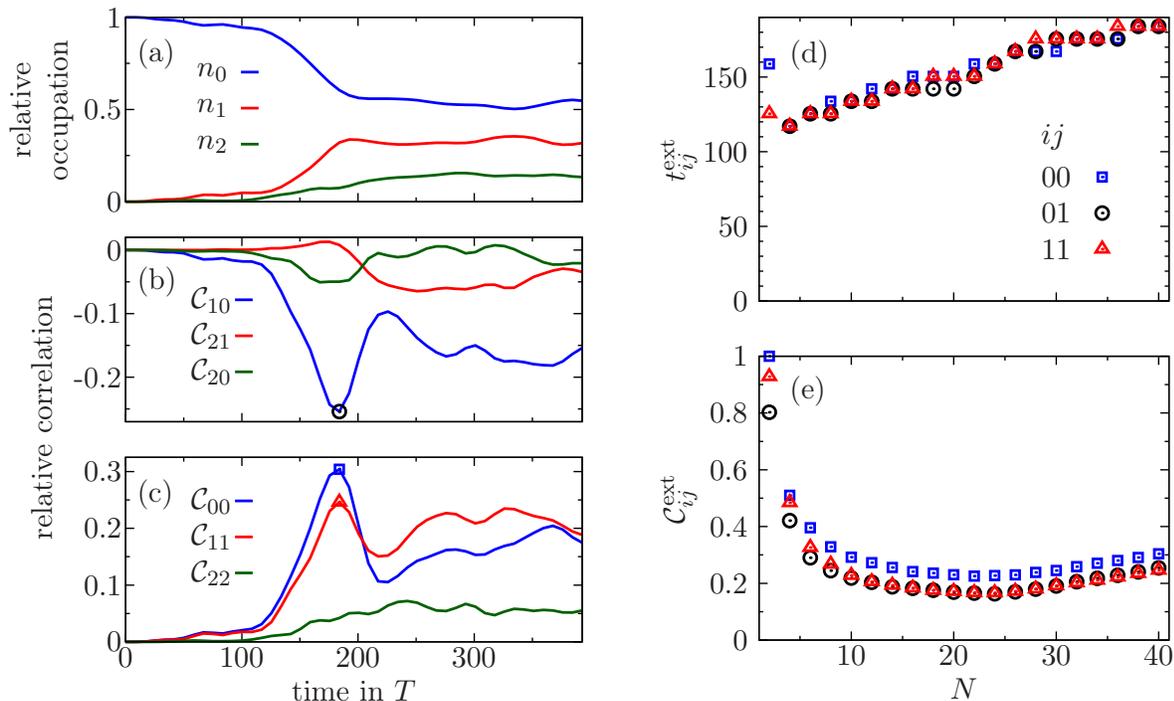}
\caption{{\it Correlated occupation in the chaotic regime.} (a)~Normalized occupation numbers $n_i\equiv\langle n_i\rangle/N$ and (b-c)~normalized particle number correlations
$\mathcal{C}_{ij}$ of the (instantaneous) natural orbitals (eigenstates of the reduced one-particle density matrix) for $N=40$ atoms. The time-point $t_{\mathrm{max}}$ of the local maximum of $\mathcal{C}_{00}$ [marked as a blue square in~(c)] for different particle numbers $N$ is given in~(d) and its value $\mathcal{C}_{00}^{\mathrm{max}}$ is shown in~(e) -- same for the maximum of $\mathcal{C}_{11}$ (red triangles) and the minimum of $\mathcal{C}_{01}$ (black circles).
In~(a-c) parameters are the same as in Fig.~\ref{fig: evolution examples}c-d.
In~(d-e), the parameters are also those of Fig.~\ref{fig: evolution examples}c-d except for the total particle number $N$, which is here plotted as a variable.}
\label{fig: chaos correlations 40}
\end{figure}
It was noted in Ref.~\cite{gardiner_nonlinear_2000} and hinted in a previous work~\cite{castin_instability_1997} that chaotic motion within GP indicates the end of validity of this approximation. This was recently reexamined in Refs.~\cite{Brezinova,BrezinovaII}. Within the 3LS, we are able to test the validity of GP directly for relatively high particle numbers. We do this by analyzing the relative occupation numbers of the natural orbitals, which are those obtained by diagonalizing the reduced one-particle density matrix, defined in a many-body context as
\begin{equation}
\rho^{(1)}_{\mu\nu}(t)\equiv\langle \hat{a}_{\mu}^{\dag}(t)\hat{a}_{\nu}(t)\rangle\,.
\end{equation}
For the two limiting cases of Rabi oscillations and self-trapping, the relative occupation of the condensate orbital (defined here as the most occupied eigenstate of $\rho^{(1)}$) stays above $98\%$ during the first $400$ driving cycles, which justifies a mean-field treatment. In contrast, for the chaotic regime, we find that the occupation of the condensate orbital decreases abruptly after some $120$ cycles (about $2$ Rabi periods) and a second natural orbital gets macroscopically occupied -- see Fig.~\ref{fig: chaos correlations 40}a. We find a low sensitivity of this behavior to the particle number.

In this chaotic regime, the macroscopic occupation of more than one orbital does not occur via a simple fragmentation of the condensate. This can be seen in the particle number correlation functions $C_{ij}\equiv\langle n_in_j\rangle-\langle n_i\rangle\langle n_j\rangle$, where $i,j$ refer to the natural orbitals.
In a fragmented condensate, $\mathcal{C}_{ij}\equiv 4C_{ij}/N^2$ would be zero (for $N\rightarrow\infty$), while for a noon (cat) type state such as e.g.
\begin{equation}\label{eqn: def: NOON state}
 |\mbox{noon}\rangle = \frac{1}{\sqrt{2}}(|N00\rangle+|0N0\rangle)
\end{equation}
its absolute value approaches unity, reflecting the creation of macroscopic particle number correlations. Figures~\ref{fig: chaos correlations 40}b-c show the time-evolution of $\mathcal{C}_{ij}$ for the same parameter set as in Figs.~\ref{fig: evolution examples}c-d. We see that, with the decrease of the condensate fraction, some particle number correlations increase in a non-negligible way. In order to check the scaling of this behavior with the particle number,
we focus on the local maximum of the variance of the condensate occupation ($\mathcal{C}_{00}$)
for different particle numbers. Figure~\ref{fig: chaos correlations 40}d shows that this maximum gets shifted to later times for higher $N$.  The value of this maximum stays above $0.2$ for all considered particle numbers and increases monotonically for $N>25$, as shown in Fig.~\ref{fig: chaos correlations 40}e. This indicates a scaling of the correlations $C_{ij}$ with $N^2$, which in turn reveals noon-like~(\ref{eqn: def: NOON state}) behavior, i.e., a dynamics dominated by a few many-body configurations differing by macroscopically large relative particle numbers. Similar observations have been made by Weiss and Teichmann~\cite{weiss_differences_2008}, investigating a driven double-well system.

Thus we find that in this chaotic regime more than one orbital gets macroscopically occupied. At the same time the particle number correlation functions scale with the square of the total particle number. This indicates a complex dynamics in which the macroscopically large particle number correlations become a substantial feature.

\section{Conclusions}
In summary, we have introduced a formalism to account for interactions in ac-driven many-body systems. In the case where only resonant states effectively intervene, we derive a Hamiltonian involving conventional operators which describes the long time dynamics of the driven system.
Although we have focused on the case where the undriven states are plane waves, the method we have developed applies to an arbitrary set of states that become resonant in the presence of ac-driving.
We have applied the method to the calculation of the coherent ratchet current carried by an asymmetrically driven atomic condensate, and compared it with continuum and truncated mean-field descriptions. We have found a rich dynamical behavior with crossovers from self-trapping to chaotic behavior to regular oscillations.
In the latter case, we find a strong departure from the mean-field picture and in particular an appreciable increase of particle number correlations.

\ack
We thank Justin Anderson and Christopher Gaul for useful discussions.
The authors acknowledge support from Spain's MINECO through Grant No. FIS2010-21372 and the Ram\'on y Cajal program (CEC), the Comunidad de Madrid through Grant Microseres, the Heidelberg Center for Quantum Dynamics, and the U.S. National Science Foundation.

\appendix
\section{Calculation of Lyapunov exponents}
\label{sec: Convergence analysis}
In this appendix we present details on the calculation of the maximal Lyapunov exponent $\lambda_{\mathrm{M}}$ given in Fig.~\ref{fig: detuning and sections}b.
The Lyapunov exponent reflects how fast the distance between two neighboring
trajectories grows over time.
We quantify the distance between two Hilbert space vectors via the Euclidean norm:
\begin{equation}\label{eqn: Euclidean distance}
 d_2(\psi_1,\psi_2)\equiv\sqrt{\langle\psi_1-\psi_2|\psi_1-\psi_2\rangle},
\end{equation}
with $|\psi_1-\psi_2\rangle\equiv|\psi_1\rangle-|\psi_2\rangle$. The maximal Lyapunov exponent for the trajectory $|\psi_0(t)\rangle$ is defined over the double limit
\begin{equation}\label{eqn: def: Lyapunov exponent}
 \lambda_{\mathrm{M}}\equiv
  \lim_{\psi_1\rightarrow\psi_0}\;
   \lim_{t\rightarrow\infty}\frac{1}{t}\;
     \log\Bigg[\frac{d(\psi_1(t),\psi_0(t))}{d(\psi_1(0),\psi_0(0))}\Bigg]\;,
\end{equation}
% Here, $\psi_1(t)$ is a solution of the GP equation with a small initial perturbation, with respect to the considered solution $\psi_0(t)$.
where $|\psi_0(t)\rangle$ and $|\psi_1(t)\rangle$ are both solutions of the GP equation.
Because of the Hamiltonian structure of the nonlinear Schr\"odinger equation, it is convenient to allow only perturbations that yield states with the same value of the Hamiltonian~\cite{Brezinova}. Note that in our effective description the Hamiltonian reflects the system's quasi-energy instead of its energy.
Within 3GP, the system's dynamics is given by the $3$-tuple $\mathbf{A}(t)=(A_+(t),A_0(t),A_-(t))$, reflecting the expansion coefficients of the condensate orbital with respect to the three modes $\pm,0$.
In order to extract the Lyapunov exponent from the system's dynamics, we consider the time evolution of the two initial states
\begin{equation}\label{eqn: perturbed states}
 \mathbf{A}_{\delta+}=\mathrm{N}_{\delta}(i\delta, 1, 0)
 \quad\mbox{and}\quad
 \mathbf{A}_{\delta-}=\mathrm{N}_{\delta}(0, 1, i\delta),
\end{equation}
where $\mathrm{N}_{\delta}=1/\sqrt{1+\delta^2}$ is the normalization factor. The parameter $\delta$ parameterizes the initial distance, $\sqrt{2}\delta$, between the trajectories $\mathbf{A}_{\delta\pm}(t)$.
The Hamiltonian of both the states~$\mathbf{A}_{\delta\pm}$ has the same value, and for~$\delta=0$, we recover the initial state considered by us.
We denote their time-dependent separation by
$d_{\delta}(t)\equiv d_2(\mathbf{A}_{\delta+}(t),\mathbf{A}_{\delta-}(t))$,
and use this quantity to estimate the Lyapunov exponent. The limit~$\psi_1\rightarrow\psi_0$ in the definition~(\ref{eqn: def: Lyapunov exponent}) translates into the limit~$\delta\rightarrow0$.

\begin{figure}[t]
 \raggedleft
  \includegraphics[width=0.99\textwidth]{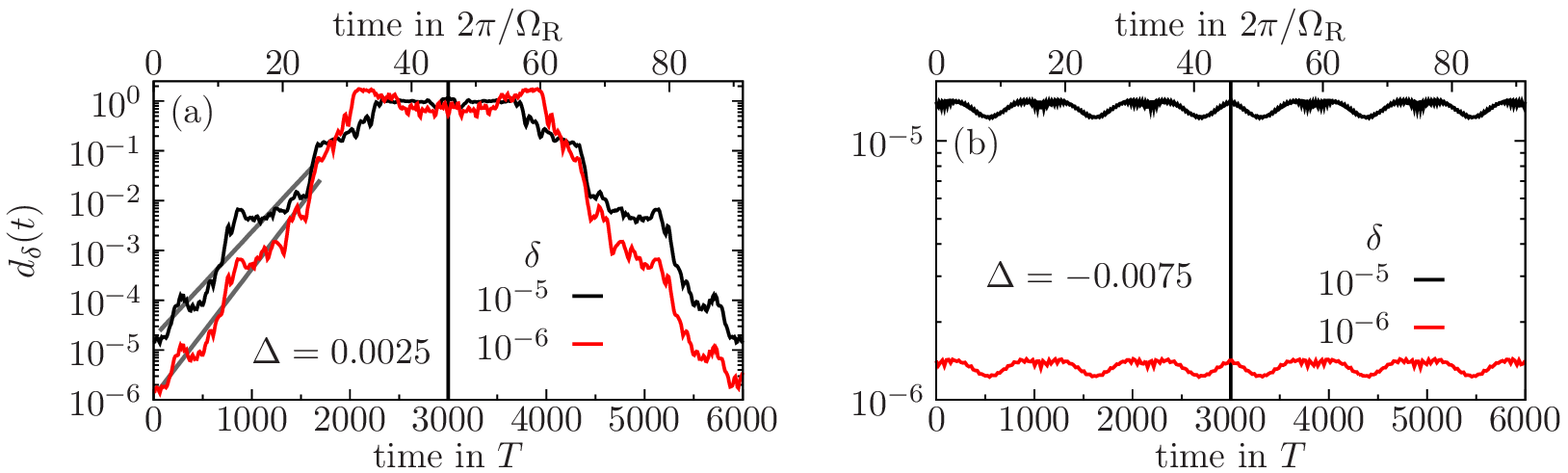}
\caption{{\it Trajectory distances over time.}
Two paradigmatic cases are considered: (a)~chaotic, and (b)~regular dynamics.
After $3000\,T$ (vertical line), the time propagation is inverted to check the accuracy.
In (a), a linear (on a log-scale) fit is included for both values of $\delta$, given by the straight gray lines.
The parameters are the same as in Figs.~\ref{fig: detuning and sections}c (b) and~\ref{fig: detuning and sections}e (a).}
 \label{fig: divergence examples}
\end{figure}

Figure~\ref{fig: divergence examples} shows the time evolution of $d_{\delta}(t)$ on a logarithmic scale for chaotic~(a), and regular dynamics~(b). Two different values of $\delta$ are considered.
For chaotic dynamics the distance between the two trajectories $d_{\delta}(t)$ grows over several orders of magnitude, until it saturates at a value around unity. This saturation occurs because the  distance $d_\delta(t)$ is bounded from above by the value $2$, since both states~$\mathbf{A}_{\delta\pm}$ are normalized to unity.
The value of $\lambda_{\mathrm{M}}$ is estimated by fitting an exponential to the rising slope of $d_\delta(t)$ before saturation is reached. For the case shown we obtain $\lambda_{\mathrm{M}}\approx0.61$ ($0.51$) for $\delta=10^{-6}$ ($10^{-5}$). The estimated values thus have a relative difference of about $18\%$, which implies a high relative uncertainty. Nevertheless, the obtained value for $\lambda_{\mathrm{M}}$ is clearly positive, indicating chaotic dynamics.

In the regular regime $d_{\delta}(t)$ deviates by less than $10\%$ of its initial value during the depicted time range, for both values of $\delta$. Chaos can be precluded when the global maximum of $d_\delta(t)$ can be decreased by choosing smaller values for $\delta$.

In all cases, the accuracy of the simulation was checked by propagating both the states~$\mathbf{A}_{\delta\pm}$ backward in time after $t=3000\,T$. The simulation is reliable, when~$d_\delta(t)$ comes back to its starting value after $t=6000\,T$. This check is necessary to ensure that the observed separation of trajectories is not due merely to numerical errors.

We additionally calculated the maximal Lyapunov exponent by a different method described e.g. in Ref.~\cite{Schuster_Deterministic}. The results obtained by that approach confirm the results presented here, by identifying the same parameter regions as chaotic (and regular) and yielding values~$\lambda_{\mathrm{M}}$ of similar magnitude.

\end{document}